\documentclass[a4paper]{jpconf}

\usepackage{graphicx}
\usepackage{sidecap}
\usepackage{amsmath}
\usepackage{wrapfigure}

\begin{document}


\title{Evidence of Josephson junction behavior in top-gated LaAlO$_3$-SrTiO$_3$}

\author{V V Bal$^1$, M M Mehta$^1$, S Ryu$^2$, H Lee$^2$, C M Folkman$^2$, C B Eom$^2$ and V Chandrasekhar$^1$}
\address{$^1$Department of Physics and Astronomy, Northwestern University, Evanston, IL 60208, USA.\\ 
 $^2$Department of Materials Science and Engineering, University of Wisconsin-Madison, Madison, WI 53706, USA.\\}

\ead{v-chandrasekhar@northwestern.edu} 
                            
\begin{abstract}
 
We demonstrate top-gate tunable Josephson junction like behavior in the two dimensional electron gas at the LaAlO$_3$-SrTiO$_3$ interface. A combination of global back-gating and local top-gating is used to define the junctions, providing an efficient way for much finer spatial control over the properties of the interface, as compared to back-gating alone. The variation of critical currents and zero bias resistances with temperature shows that the junctions behave like short, overdamped weak links.  This technique could be an important tool to illuminate the nature of superconductivity in the LaAlO$_3$-SrTiO$_3$ interface system. 
\end{abstract}

The two dimensional electron gas (2DEG) at the interface of the insulating oxides LaAlO$_3$ (LAO) and SrTiO$_3$ (STO), has attracted a lot of attention since its discovery in 2004 \cite{ohtomo,caviglia,thiel}. It was found that the 2DEG undergoes a superconducting transition below about 300 mK \cite{reyren}. An important question is the nature of superconductivity, as it occurs in two dimensions, in an environment of strong spin-orbit coupling \cite{gariglio}, and in coexistence with magnetism \cite{dikin, bert, li}. Various theories have been proposed to explain superconductivity in this system, including a Fulde-Ferrel-Larkin-Ovchinnikov state \cite{lee}, a mixed parity state \cite{hu}, and a conventional superconducting state involving electrons from $d_{xy}$ orbitals in a few TiO$_2$ layers of STO next to the interface
\cite{banerjee}. Penetration depth experiments \cite{bert2} point to doubly gapped \textit{s}-wave superconductivity with strong coupling, whereas out-of-plane tunneling measurements do not rule out other symmetries \cite{richter}. Thus a clear consensus is still lacking. Phase sensitive tests involving Josephson junctions (JJ) are an important tool to garner information about the order parameter symmetry of a superconductor \cite{van harlingen,golubov}. In this work, we demonstrate Josephson junction like behavior in the 2DEG at the LAO-STO interface.

It is known that the 2DEG at the LAO-STO interface can be tuned through a superconductor-insulator transition (SIT) by the application of a back-gate \cite{caviglia} or top-gate voltage \cite{forg,eerkes}. The strength of superconductivity is a nonmonotonic function of the gate voltage. This offers a handy way to define tunable barriers in this system, with fine spatial control, using a combination of local top-gates and a global back-gate. We expect that by applying appropriate top-gate voltages, we can tune the area underneath the top-gate from a superconductor, through a normal metal, to finally an insulator.

We measured two devices, both of which showed similar behavior, corresponding to short, overdamped junctions. In the remainder of the paper, we focus on one of these devices. Figure 1 shows a scanning electron micrograph (SEM) of the device, which was fabricated on 10 unit cells of LAO grown epitaxially by pulsed laser deposition (PLD) on TiO$_2$ terminated (001) STO. The details of the PLD synthesis are discussed in earlier papers \cite{park, bark}. Using photolithography and Ar ion milling, Hall bars of width 100 $\mu$m were patterned onto the sample. Earlier measurements showed that deposition of the top-gate electrode directly on top of LAO can lead to shorting of the top-gate and the 2DEG after just a few sweeps of the top-gate voltage, $V_{tg}$, even though LAO itself is an insulator. Hence we used e-beam lithography and e-beam evaporation of 70 nm of SiO$_2$ in an oxygen atmosphere to define an additional insulating layer, covering one of the sections of the Hall bar between two pairs of voltage probes. This was done in order to have a more reliable barrier between the top-gate and 2DEG, so that we could use a larger range of values of $V_{tg}$ and hence go deeper into the insulating regime of the 2DEG. Finally, the same lithography techniques were used to deposit Ti/Au for the top-gate on top of the SiO$_2$. The top-gated region in its narrowest section was a strip of length 6 ${\mu}$m and width 180 nm. For certain values of $V_{tg}$ and temperature, $T$, we expect the area underneath the top-gate to be non-superconducting, and the 180 nm wide section to be the least resistance path for the current, as we discuss below. If the superconducting correlations of the two non-top-gated superconducting banks on either side of the top-gated region can be supported across this 180 nm wide section, then all current would flow only through it, thus defining a weak link, or a JJ, between the banks. 

\begin{SCfigure}
      \centering
      \includegraphics[width=0.6\textwidth,height=5cm]{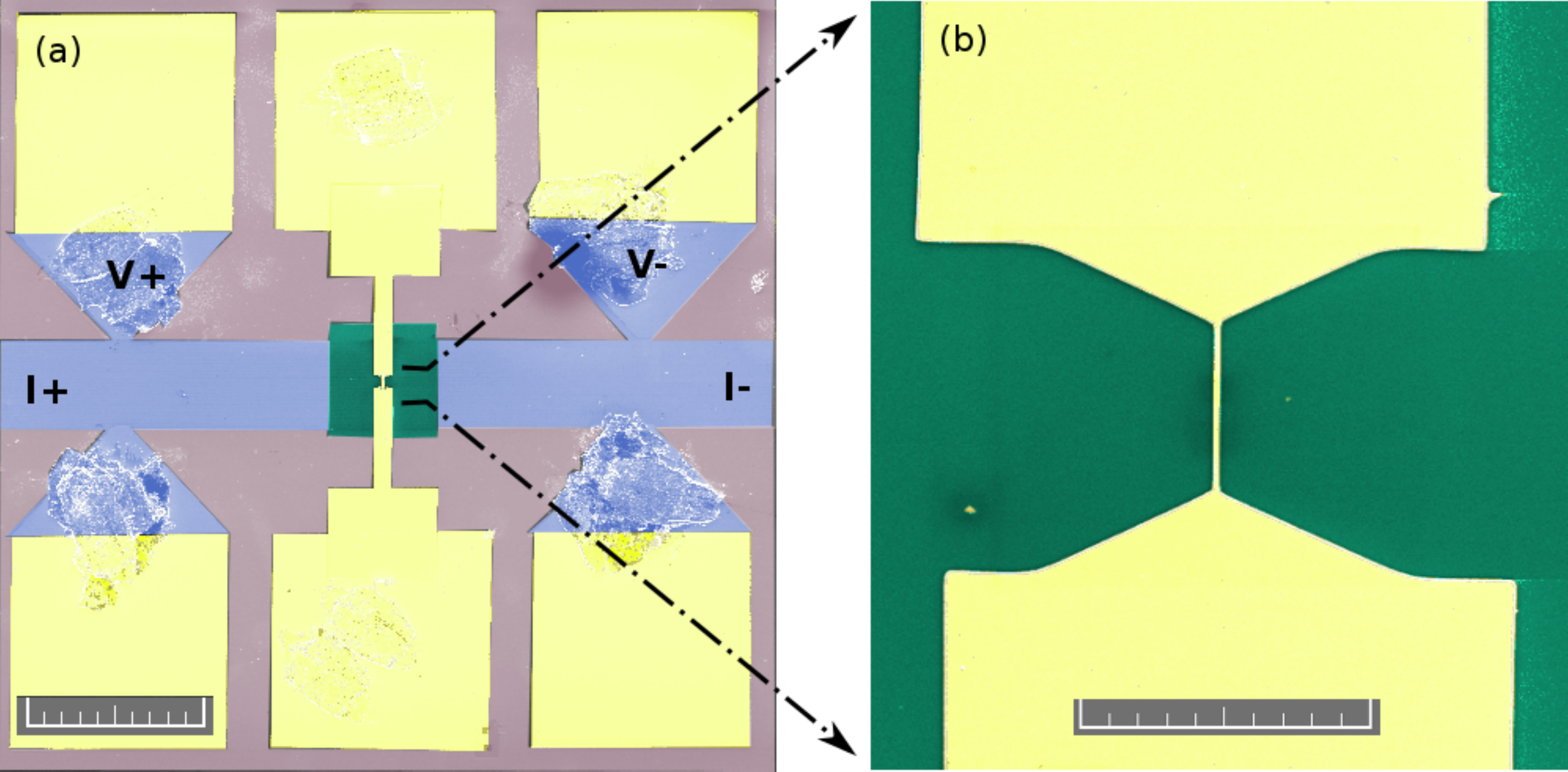}
      \caption{\textbf{a} Colorized SEM of part of the Hall bar. The blue area is the LAO-STO. The green area is the SiO$_2$ layer covering the LAO-STO. The gold area is Ti/Au for the leads and the top-gate. It also covers the triangular voltage probes of the Hall bar. Residue from wire bonds can be seen in the image. The scale bar is 200 $\mu$m. \textbf{b} Colorized SEM of the narrowest part of the device. The scale bar is 10 $\mu$m.}
\end{SCfigure}

Measurements were performed using a standard lockin amplifier technique with a SRS124 lockin amplifier and a home built current source, with an ac excitation frequency of 107 Hz and amplitude $\sim$ 10 nA. An Agilent synthesizer and a home-built summer were used to apply a dc current bias. The sample was measured in an Oxford MX100 dilution refrigerator with a base temperature of 25 mK. For all top-gate voltages studied, the leakage current from top-gate electrode to the 2DEG was less than 14 nA, which is a negligible dc offset as compared to critical current values of the device.

$dV/dI$ vs. $I_{dc}$ at $V_{tg}$ = 0 V and $V_{bg}$ = 0 V, before sweeping $V_{tg}$, is shown by the red curve in Fig. 2a. $dV/dI$ was found to be hysteretic in $I_{dc}$, with normal state resistance $R_N$ = 4 k$\Omega$. We then swept $V_{tg}$ from 0V $\to$ +40 V $\to$ -40 V $\to$ 0 V, with $V_{bg}$ = 0 V. As we swept to high negative values of $V_{tg}$, the $dV/dI$ increased drastically. The $dV/dI$ vs. $I_{dc}$ characteristics after this $V_{tg}$ sweep, at $V_{tg}$ = 0 V and $V_{bg}$ = 0 V were very different from those before the $V_{tg}$ sweep, and are shown by the black curve in Fig. 2a. We note four characteristic differences. First, $R_N$ increased to 11.8 k$\Omega$, which is much higher than its value before sweeping $V_{tg}$. Second, there are two distinct current scales in the data after the $V_{tg}$ sweep. One, which we denote by $I_{ci}$, is where the system transitions out of the zero resistance state, and the other, which we denote by $I_{co}$, is beyond which the superconducting banks themselves go normal. We define $I_{ci}$ as the current at which $dV/dI$ rises beyond 500 $\Omega$, and $I_{co}$ as the position of the most prominent peak in $dV/dI$. Third, additional structure appears in $dV/dI$, characteristic of the particular $V_{tg}$ and reproducible in different $I_{dc}$ sweeps. Finally, the $dV/dI$ vs. $I_{dc}$ plot has negligible hysteresis near $I_{ci}$. We now discuss each of these features in detail.

Critical current measurements before sweeping $V_{tg}$ showed that superconductivity for our sample is strongest at $V_{bg}$ = 90 V. In order to ensure that the banks are strongly superconducting, we set $V_{bg}$ = 90 V for all further measurements. Transport characteristics of the device were found to be hysteretic in $V_{tg}$, hence all measurements were made going from positive to negative values of $V_{tg}$. 

\begin{SCfigure}
      \centering
      \includegraphics[width=0.65\textwidth,height=5.5cm]{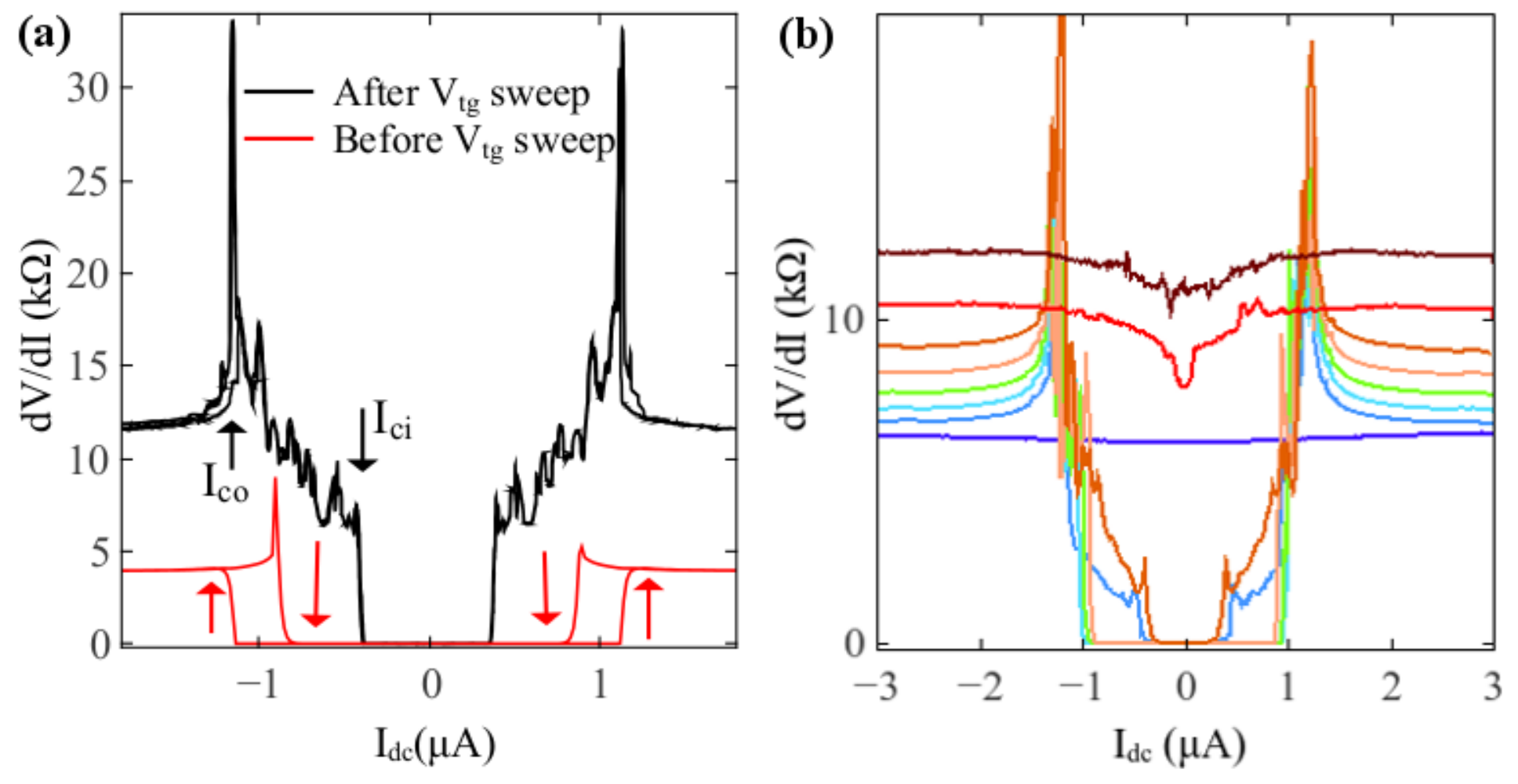}
      \caption{\textbf{a} $dV/dI$ vs. $I_{dc}$ for the device, after sweeping $V_{tg}$, at 30 mK, with $V_{tg}$, $V_{bg}$ = 0 V. Both sweep directions are shown. \textbf{b}  $dV/dI$ vs. $I_{dc}$ at 30 mK, at $V_{bg}$ = 90 V, for 8 different values of $V_{tg}$. Going from the curve with the smallest $R_N$ (blue curve) to the largest $R_N$ (brown curve), $V_{tg}$ decreases from +30 V to -40 V, in steps of 10 V. Only one sweep direction is shown.}
     \end{SCfigure}

Figure 2b shows how $dV/dI$ varies with $I_{dc}$ for various values of $V_{tg}$, at $T$ = 30 mK. We observe that $R_{N}$ increases monotonically with decreasing $V_{tg}$. $R_{N}$ is a sum of the normal state resistance of the banks, which is independent of $V_{tg}$, and the normal state resistance of the top-gated part, which is a function of $V_{tg}$. Before the $V_{tg}$ sweep, at $V_{bg}$ = 90 V, $R_N$ of the 600 $\mu$m long and 100 $\mu$m wide section of the Hall bar was 3 k$\Omega$, which gives a sheet resistance of the banks, $R_{NB}^s$, of 500 $\Omega$. The top-gated section can be considered to be a parallel combination of the resistances of the two broader top-gated parts (20 $\mu$m long and 40 $\mu$m wide), which correspond to 0.5 squares each, and the narrowest top-gated part, which corresponds to 0.03 squares. Knowing $R_{NB}^s$ at $V_{bg}$ = 90 V and assuming that the sheet resistance is uniform throughout the top-gated part, the normal state sheet resistance $R_{NG}^s$ of the top-gated part can be estimated. $R_{NG}^s$ increases from 220 k$\Omega$ to 430 k$\Omega$ in going from $V_{tg}$ = 30 V to -40 V. These values of $R_{NG}^s$ are two orders of magnitude greater than the maximum values obtained by back-gating \cite{dikin}, which demonstrates the efficiency of the top-gate.

We now discuss the structure in $dV/dI$ vs. $I_{dc}$ and the origin of the two current scales, $I_{ci}$ and $I_{co}$. Since we back-gated our sample at 90 V to be maximally superconducting, and from the fact that the strength of superconductivity is a non-monotonic function of gate voltage \cite{caviglia}, we know that the top-gated region cannot be a stronger superconductor than the banks, which means it is either a normal metal, an insulator, or a weaker superconductor. For values of $V_{tg}$ at which the top-gated part is a normal metal or an insulator, the narrowest 180 nm wide section of the top-gated region is the least resistance path for the current for all temperatures. This is also true for values of $V_{tg}$ at which the top-gated region is a weaker superconductor than the banks, for the temperature regime below the critical temperature of the banks but above the critical temperature of the top-gated region.

In these cases, the superconducting correlations of the banks cannot be supported over the width of the broad top-gated part (20 microns). The 180 nm wide section of the top-gated region, however, may support superconductivity across its width, hence the supercurrent is 
\begin{wrapfigure}{r}{0.48\textwidth}
 \centering
  \includegraphics[width=0.41\textwidth]{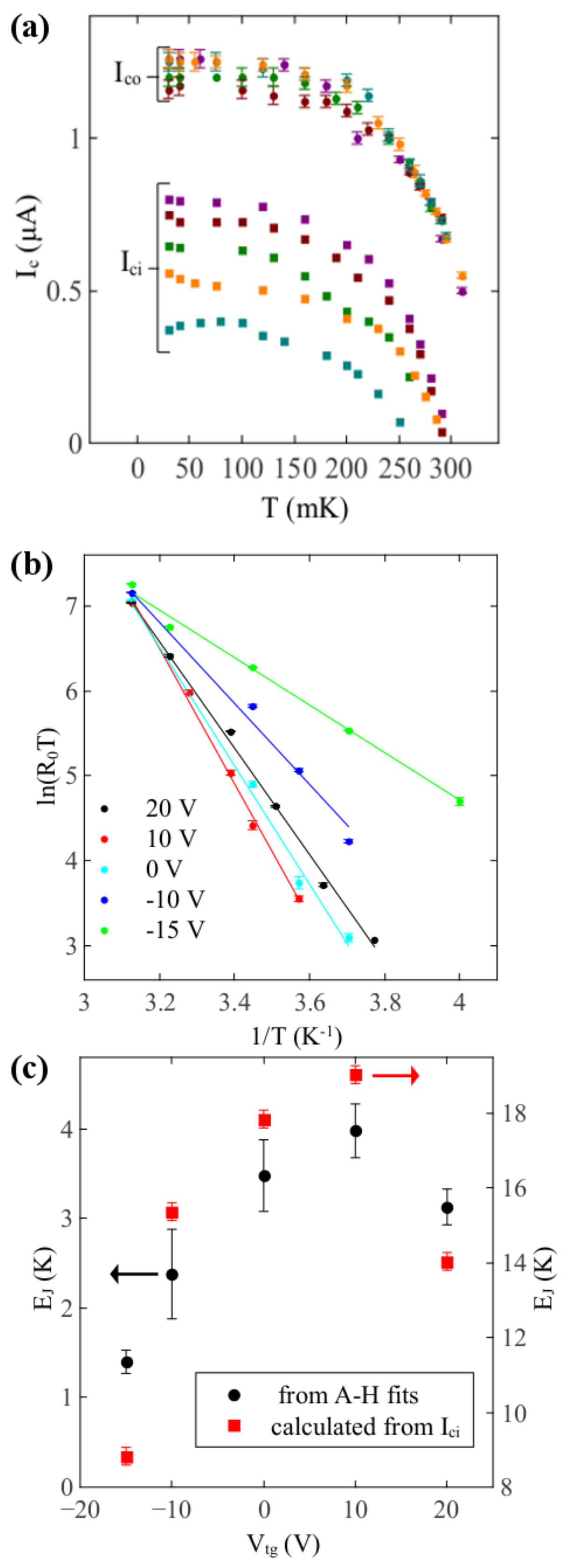}
   \small\caption{\textbf{a} $I_{ci}$ (squares) and $I_{co}$ (circles) vs. $T$ for various $V_{tg}$. Orange denotes data for 20 V, purple for 10 V, brown for 0 V, green for -10 V and blue for -15 V. \textbf{b} Fits to the Ambegaokar-Halperin model. \textbf{c} $E_J$ vs. $V_{tg}$ as calculated from fits to the Ambegaokar-Halperin model, and independently from measured $I_{ci}$s.}
\end{wrapfigure}
confined to flow only through this section. This then defines a JJ between the non-top-gated superconducting banks. The zero resistance state would disappear when $I_{dc}$ exceeds the critical current of this narrow section, which corresponds to $I_{ci}$. Once $I_{dc}$ exceeds the critical current of the superconducting banks, the final high-bias resistance, which we call $R_N$, is reached. From Fig. 2b, we see that the $I_{co}$ peaks for various $V_{tg}$ fall on top of each other, supporting the claim that they are related to the superconducting banks themselves going normal. The $I_{ci}$ values on the other hand vary strongly with $V_{tg}$. The electric field due to the top-gate extends over some region beyond the top-gate electrode. We conjecture that the peaks in $dV/dI$ between $I_{ci}$ and $I_{co}$ appear as different parts of this region attain their critical currents. Hence these features are characteristic of the specific value of $V_{tg}$.

For values of $V_{tg}$ at which the top-gated region is a weaker superconductor than the banks, for the temperature regime below the critical temperature of the top-gated region, however, the entire 100 $\mu$m cross-section of the Hall bar is superconducting. In this regime, we do not have current concentration through the narrowest section of the top-gated region, and hence no junction behavior. From our data, although we cannot be sure about which regime we are in for a given value of $V_{tg}$, it is clear that at least for some values of $V_{tg}$ and $T$, we have a weak link which is tunable by the top-gate.

In order to categorize our weak link, we measured $dV/dI$ as a function of $I_{dc}$ at different temperatures for various $V_{tg}$. Figure 3a shows the variation of $I_{ci}$ and $I_{co}$ with $T$ for five different values of $V_{tg}$. The behavior of the two data sets is distinctly different, as the variation of $I_{co}$ with $T$ seems to be almost independent of $V_{tg}$, while the $I_{ci}$ vs. $T$ plots are very different for different $V_{tg}$. For long junctions, $I_{ci}$ should decay exponentially with $T$. No exponential decay of $I_{ci}$ is evident from Fig. 3b, at least for the $V_{tg}$ tested, and hence we conjecture that we are in the short junction limit, although we cannot be sure what type of barrier separates the banks.

Near $T_C$, the current-phase relation of JJs becomes approximately sinusoidal \cite{golubov}, and hence the junction dynamics can be described by the well known tilted washboard model \cite{tinkham}. JJs can be classified as either underdamped or overdamped based on junction dynamics, irrespective of the kind of barrier between the banks. The $I-V$ characteristics of our device are not hysteretic, and hence suggest an overdamped junction \cite{tinkham}. We attempted to fit the zero bias resistance, $R_0$, to the Ambegaokar-Halperin thermally activated phase slip model for overdamped JJs \cite{ambegaokar, halperin}. According to this model, $R_0 \propto 1/T \mathrm{exp}(-2E_J / k_B T)$, where $E_J$ is the Josephson coupling energy. The proportionality holds if $E_J >> k_BT$. In our device, the estimated $E_J$s from the fits are about an order of magnitude larger than $k_BT$, ranging from 1.4 eV for $V_{tg}$ = -15 V to 3.98 eV for $V_{tg}$ = 10 V. The linear fits of ln$(R_0 T)$ vs. $1/T$ are shown for various $V_{tg}$ in Fig. 3b. From the slopes of the linear fits we extract $E_J$, which are plotted as a function of $V_{tg}$ in Fig. 3c, again showing a non-monotonic dependence on $V_{tg}$. On the same plot we also show $E_J$ obtained from the measured $I_{ci}$, as calculated from $E_J = \hbar I_{ci} / 2 e$. We see that they too show the expected non-monotonic dependence on $V_{tg}$, although the $E_J$ values obtained in this way are larger than those obtained from the fits. We cannot rule out the possibility that some parts of the broader top-gated sections could also contribute to transport, leading to inaccuracy in the measured $I_{ci}$. However, the fact that the transport parameters vary systematically with $V_{tg}$, suggests that $V_{tg}$ provides a very good handle on the weak link properties.

In conclusion, we have demonstrated top-gate tunable weak links in the LAO-STO system. The weak links fit reasonably well to an overdamped junction model. The strength of the superconducting order parameter is a nonmonotonic function of the top-gate voltage, whereas the normal state resistance is a monotonic function of the top gate voltage, similar to what has been reported for back-gate tuning. In order to better characterize the junctions, we need to measure devices with blanket top gates, enabling us to have better control over device design and hence a better understanding of the system.

The U.S. Department of Energy, Office of Basic Energy Sciences supported the work at Northwestern University through Grant No. DE-FG02-06ER46346. Research at UW-Madison was supported by the US Department of Energy, Office of Basic Energy Sciences, Division of Materials Sciences and Engineering under Award number  DE-FG02-06ER46327.

\end{document}